\documentclass[abstract]{aa}
\usepackage[lofdepth,lotdepth]{subfig}
\usepackage{graphicx}
\usepackage{txfonts}
\usepackage{natbib}
\usepackage{float}

\newcommand{\mJybeam}{mJy beam$^{-1}$}

\newcommand{\kms}{$\,$km$\,$s$^{-1}$}

\newcommand{\kmsmpc}{$\,$km$\,$s$^{-1}$Mpc$^{-1}$}

\bibpunct{(}{)}{;}{a}{}{,}

\defcitealias{Planck2016}{Planck Collaboration XIII}

\begin{document} 

\title{LOFAR first look at the giant radio galaxy 3C~236}

\titlerunning{LOFAR studies of 3C~236}
\authorrunning{Shulevski et al.}	
\author{A.~Shulevski
	\inst{1,2}
	\and
	P.~D.~Barthel
	\inst{3}
	\and
	R.~Morganti
	\inst{2,3}
	\and
    J.~J.~Harwood
	\inst{4}
	\and
	M.~Brienza
	\inst{5}
	\and
	T.~W.~Shimwell
	\inst{1,6}
	\and
    H.~J.~A.~R\"{o}ttgering
    \inst{6}
    \and
	G.~J.~White
	\inst{7,8}
	\and
	J.~R.~Callingham
	\inst{2}
	\and
	S.~Mooney
	\inst{9}
    \and
    D.~A.~Rafferty
    \inst{10}
	}
	\institute{
	Anton Pannekoek Institute for Astronomy, University of Amsterdam, Postbus 94249, 1090 GE Amsterdam, The Netherlands\\
	\email{a.shulevski@uva.nl} 
	\and
	ASTRON, the Netherlands Institute for Radio Astronomy, Postbus 2, 7990 AA, Dwingeloo, The Netherlands
	\and
	University of Groningen, Kapteyn Astronomical Institute, Landleven 12, 9747 AD Groningen, The Netherlands
	\and
    Centre for Astrophysics Research, School of Physics, Astronomy and Mathematics, University of Hertfordshire, College Lane, Hatfield AL10 9AB, UK 
	\and
	INAF – Istituto di Radioastronomia, via Gobetti 101, 40129, Bologna, Italy
	\and
    Leiden Observatory, Leiden University, PO Box 9513, NL-2300 RA Leiden, the Netherlands
    \and
	School of Physical Sciences, The Open University, Walton Hall, Milton Keynes, MK7 6AA, UK
	\and
	Space Science \& Technology Department, The Rutherford Appleton Laboratory, Chilton, Didcot, Oxfordshire OX11 0NL, UK
	\and
	School of Physics, University College Dublin, Belfield, Dublin 4, Republic of Ireland
    \and
    Hamburger Sternwarte, University of Hamburg, Gojenbergsweg 112, 21029 Hamburg, Germany	
	}
\date{\today}
 
\abstract
{We have examined the giant radio galaxy 3C~236 using LOFAR at 143 MHz down to an angular resolution of $ 7\arcsec $, in combination with observations at higher frequencies. We have used the low frequency data to derive spectral index maps with the highest resolution yet at these low frequencies. We confirm a previous detection of an inner hotspot in the north-west lobe and for the first time observe that the south-east lobe hotspot is in fact a triple hotspot, which may point to an intermittent source activity. Also, the spectral index map of 3C~236 shows that the spectral steepening at the inner region of the northern lobe is prominent at low frequencies. The outer regions of both lobes show spectral flattening, in contrast with previous high frequency studies. We derive spectral age estimates for the lobes, as well as particle densities of the IGM at various locations. We propose that the morphological differences between the lobes are driven by variations in the ambient medium density as well as the source activity history.}


\maketitle

\section{Introduction}
\label{intro}

Giant radio galaxies (GRGs) are radio galaxies whose radio emitting regions (jets, lobes) extend over projected distances $ \geq 1 $ Mpc \citep{RefWorks:222, RefWorks:40, RefWorks:254, RefWorks:236, RefWorks:101}. Their morphology can be classified as core-brightened FRI or edge-brightened FRII \citep{Fanaroff1974}. There is no evidence that they are particularly more energetic than the general population of radio galaxies \citep[e.g.][]{Lara2001}. 

A low-density environment may be the key factor enabling their growth to such large sizes. \cite{RefWorks:148} have indeed found that the surrounding medium for their sample of GRGs is an order of magnitude less dense than that around smaller radio sources. Hence the radio sources can expand freely, implying that expansion losses rather than radiative losses are the dominant energy loss mechanism for the relativistic particle populations in the radio lobes.

Apart from their size, GRGs are not fundamentally different from other radio galaxies, and they are expected to be subject to the same processes that are present in smaller radio galaxies. The AGN that power them go through a cycle of active and inactive periods \citep[e.g.][]{McNamara2007, Morganti2017}. Hence, we might expect GRGs to show evidence of multiple activity episodes, both in their radio morphology and spectra. They may exhibit double-double morphologies \citep[some examples can be found in the sample of][]{Malarecki2013} and show signs of spectral curvature indicating radiative ageing of the relativistic particles responsible for their extended radio emission.

There are several studies of the ages of GRGs using radio data. \cite{RefWorks:148} have performed radiative ageing analysis of five giant radio galaxies (including 3C~236), obtaining ages less than 400 Myr. More recently \cite{Hunik2016} have presented a restarted giant radio galaxy for which they derive a radiative age of 160 Myr. Also, Cantwell et al. (submitted) studied NGC~6251 and found ages raging from 50 Myr to greater than 200 Myr. \cite{Orru2015} have studied the famous double-double GRG B1864+620 and showed that the source ages derived for its outer pair of radio lobes indicate that the AGN activity has stopped in the recent past. 

For many years following its discovery in the late 1950s, 3C~236 was an unresolved source. It was catalogued as such in the first 3C catalogue and kept its status up to and including the study of \cite{RefWorks:224}. However, using the Westerbork Synthesis Radio Telescope (WSRT), \cite{RefWorks:222} discovered low surface brightness radio lobes emanating from the compact source, extending over a total projected linear size 4.5 Mpc ($ z = 0.1005 $)\footnote{We assume a flat $\Lambda$CDM cosmology with $ H_{0} $ = 67.8\kmsmpc and $ \Omega_{m} $ = 0.308, taken from the  cosmological results of the full-mission Planck observations \citepalias{Planck2016}, and use the cosmology calculator of \cite{Wright2006}. At the redshift of 3C~236, 1$\arcsec$ = 1.8 kpc.}. For decades, it was the largest known radio galaxy \citep[but see][for the current record holder]{RefWorks:236}, hence 3C~236 is listed as a GRG in radio survey catalogues. 

\cite{RefWorks:130} investigated the radio morphology at a variety of scales. They noted that the low surface brightness emission of the lobes, especially the north-west (NW) one, shows a large-scale (300 kpc) wiggle, possibly associated with the jet slightly changing its orientation over the source's lifetime (see their Figure 4). The NW lobe terminates in a diffuse plateau, and there is an inner hotspot embedded in it, which may indicate a separate episode of AGN activity or intermittent accretion. The south-east (SE) lobe is more extended and shows a double hotspot morphology which the authors suggest may be caused by an oblique shock deflecting the jet. \cite{RefWorks:225} studied the spectral index variations across the lobes and found that the spectral index steepens going from the outer edges of the lobes towards the host galaxy, similar with that observed in (hotspot dominated) FRII radio galaxies.

The host galaxy of 3C~236 has been studied in detail by \citet{RefWorks:227}, \citet{RefWorks:228} and \citet{RefWorks:51}. Hubble Space Telescope (HST) imaging has revealed repeated bursts of star formation (on timescales of $ \sim 10^{7} $ and $ \sim 10^{9} $ yrs) in a warped dusty disk surrounding the AGN. This suggests that the younger starburst may be connected to the AGN reactivation which produced the currently active Compact Steep Spectrum (CSS) radio source at its centre \citep{RefWorks:51}. Thus, 3C~236 is an example of a radio galaxy showing signs of multiple epochs of radio activity.

The central regions of this radio galaxy are rich in gas. Atomic neutral hydrogen was revealed as a deep narrow absorption feature near the systemic velocity by \cite{vanGorkom1989}. The distribution of this gas was studied at high spatial resolution using VLBI techniques by \cite{Struve2012}, who speculate about the radio source interacting with the cold ISM gas. \cite{Morganti2005} have discovered a broad and shallow wing, blueshifted up to 1000 \kms . This absorption is associated with a fast outflow (a feature shared by a number of restarted radio galaxies), and has been recently traced also on VLBI (pc) scales \citep{Schulz2018}. The presence of cold molecular gas (traced by CO) was found by \cite{Labiano2013}, using the Plateau de Bure Interferometer at 209.5 GHz. The gas appeared to be rotating around the AGN, and was observed to be co-spatial with the dusty disk in which the AGN is embedded.

With the advent of the LOw Frequency ARray \citep[LOFAR;][]{RefWorks:157} it is now possible, for the first time, to study the extended GRG morphology in a comprehensive manner at very low frequencies. LOFAR is sensitive to extended low surface brightness features due to its dense sampling of short spacings in the UV plane, while at the same time enabling high spatial resolution imaging leveraging its long (Dutch) baselines.

Within the framework of the LOFAR Surveys Key Science Project, the nearby-AGN working group has observed two GRGs: 3C~236 and NGC~6251. These are among the largest GRGs and have never been studied in such detail as LOFAR can provide in its frequency range. In this work, we present the results related to 3C~236. Our goal is to perform high resolution mapping of the radio morphology of its extended structure at the lowest frequencies to date, enabling us to trace the oldest emission regions. Our aim is also to extend the (resolved) spectral index studies of this object a factor of two lower in frequency compared to previous studies. This enables us to place tighter constraints on the source energetics and activity history, tying in with previous studies of this object.

The organization of this work is as follows. Section \ref{data} describes the observations and the reduction procedure. In Section \ref{res} we outline our results, we discuss them in Section \ref{dis} and conclude with Section \ref{con}.

\section{Observations}
\label{data}

\subsection{LOFAR observations}

The observations were performed with the LOFAR telescope operating in high-band (HBA) mode, for a total on-source time of 8 hours, on the morning of October 9, 2018. Details of the observation are given in Table \ref{table:obs}. Main and secondary calibrator scans were also performed before and after the target observation, ten minutes each in duration.

\begin{table}[!htpb]
\noindent \caption{\small LOFAR configuration}
\label{table:obs}
\centering
\small
\begin{tabular}{ll}
\hline\hline\\
Project code & LT10\_010 \\
Central Frequency [MHz] & 143.65 \\
Bandwidth [MHz] & 47 \\
Integration time & 1 second \\
Observation duration & 8 hours\\
Polarization & Full Stokes \\
Primary flux reference & 3C~147\\
\hline
\end{tabular}
\end{table}

The data were initially pre-processed (flagged and averaged) by the LOFAR observatory pipeline \citep{RefWorks:180}. The station gains were determined using the main calibrator pointing and set to the \citet{RefWorks:181} flux density scale.

\begin{figure*}[!ht]
\centering
\includegraphics[width=0.45\textwidth]{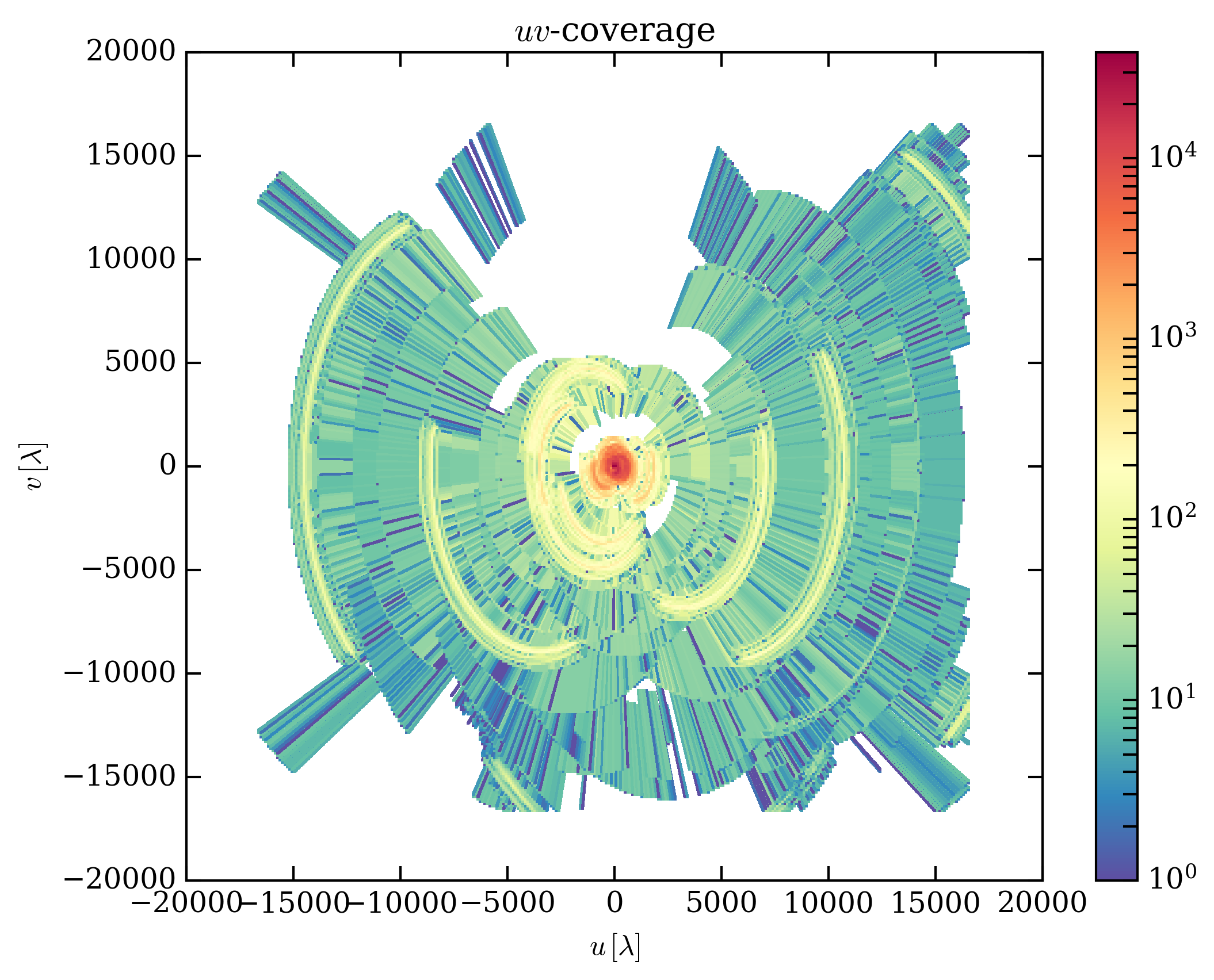}
\includegraphics[width=0.45\textwidth]{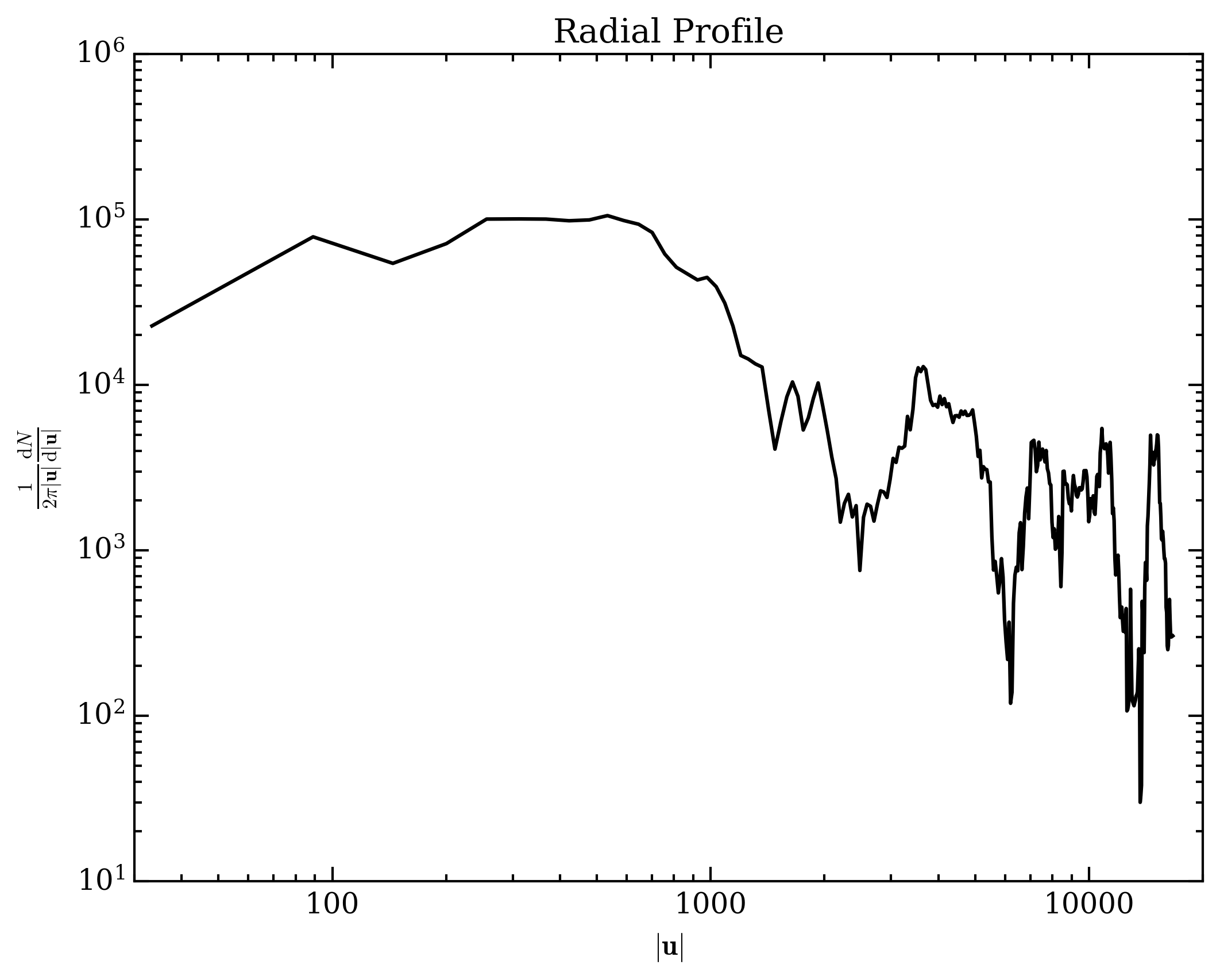}
\caption{UV coverage (left) and its radially averaged profile of the LOFAR data.}
\label{3C236:uv}
\end{figure*}

To image the entire field of view at these low frequencies, the influence of the ionosphere has to be properly taken into account. The observation used was part of the ongoing LOFAR Two-metre Sky Survey (LoTSS) project and the data were processed using its reduction pipelines which perform directional (self) calibration and imaging. For a full description, please refer to \cite{Shimwell2017,Shimwell2019}.

3C~236 was the brightest source in the field, and our main source of interest, so we did not calibrate and image across the entire FoV, focusing only on the area around the target (Tasse et al., van Weeren et al. in prep.). The calibrated data set (with uv-coverage as shown in Figure \ref{3C236:uv}) was imaged with WSclean \citep{Offringa2014} using multi-scale cleaning; scales of $ 0 - 2^{n}\, , n = [2, 6] $ pixels. The image shown in the main panel of Figure \ref{3C236:map} was obtained using a UV taper of $ 7.4\,\mathrm{k}\lambda $ and Briggs weights with robustness set to $ -0.5 $. To emphasize source structure on the smallest scale, we have imaged without tapering using the same weights as described previously. The final image properties are listed in Table \ref{table:imgs}.

LOFAR flux densities are known to suffer from a systematic effect when the amplitude scale is set by transferring the gains from calibrator to target pointing. Different elevations of the target and calibrator sources will yield different gain normalization of the LOFAR beam pattern, which can appear as a frequency dependent flux density offset \citep{Shimwell2019}. To further verify our flux scale as measured from the images we have obtained, a check was performed measuring the flux density of the unresolved bright core as well as another nearby point source and comparing it with catalogue values; we found a residual flux excess of 42\% which we corrected for by down-scaling the LOFAR image.

\subsection{Literature data}
 
In order to perform the spectral study  of 3C~236, we have collected images from the literature that trace the emission of the extended lobes and could be combined with the LOFAR data. This combination needs to be done with care and in Section \ref{spec_ind} we comment more on the caveats.

We have used legacy Very Large Array (VLA) survey (NVSS\footnote{NVSS stands for the NRAO VLA Sky Survey carried out at a frequency of 1.4 GHz \citep{RefWorks:139}}), as well as Westerbork Synthesis Radio Telescope (WSRT) images. The image properties are listed in Table \ref{table:imgs}. The mid and high resolution LOFAR images are shown in Figure \ref{3C236:map}. The images collected allows us to produce spectral index maps and derive the spectral curvature and ages of the lobes.

\begin{figure}[!ht]
\centering
\includegraphics[width=0.5\textwidth]{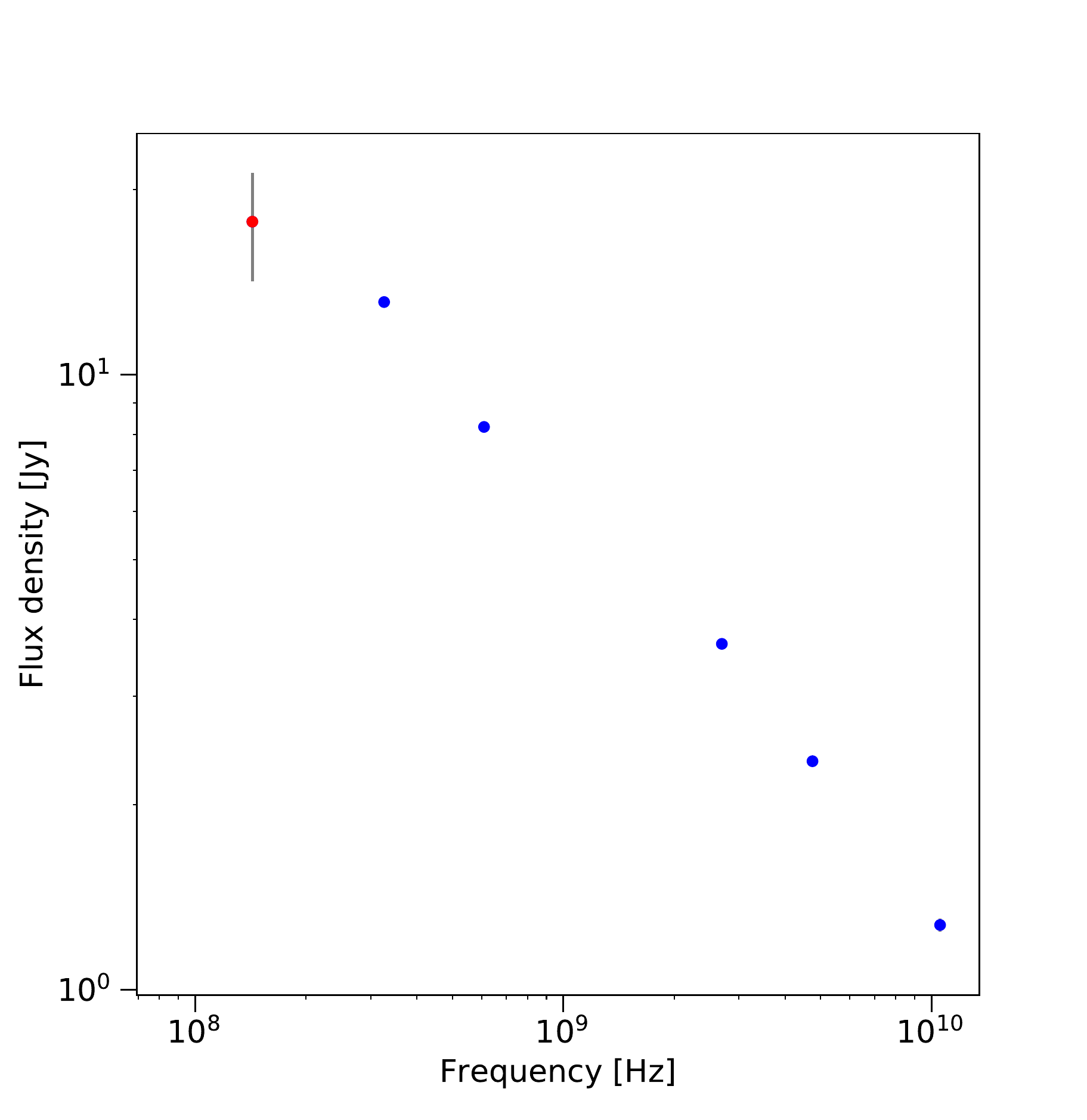}
\caption{Integrated flux density of 3C~236.}
\label{3C236:int}
\end{figure}

In Figure \ref{3C236:int} we plot the integrated source flux density measurements taken from the study of \cite{Mack1997} (with frequency coverage up to 10550 MHz, given in Table \ref{table:intflux}), together with those measured in our low resolution LOFAR map and the NVSS map, both listed in Table \ref{table:imgs}.

The LOFAR integrated flux density (marked in red) shows a slight flattening of the integrated source spectrum at low frequencies compared to the points at high frequency. This is to be expected, as we shall discuss in the forthcoming sections of this work. This flattening was hinted at in previous studies \citep[e.g.][]{Mack1997}. As can be discerned from Figure \ref{3C236:int}, the flux density scale of our LOFAR observations is as expected, within the errors.

\begin{table}
\centering
\noindent \caption{\small Image properties.}
\label{table:imgs}
\small
\begin{tabular}{c c c c c}
\hline\hline\\
\small Instrument & \small $ \nu $ [MHz] & \small $ \Delta \nu $ [MHz] & \small $ \sigma $ [mJy/b] & \small Beam size\\
\hline\\
LOFAR & 143.65 & 46.87 & 0.26 & $ 11.77\arcsec \times 6.82\arcsec $ \\
LOFAR & 143.65 & 46.87 & 0.5 & $ 23.81\arcsec \times 19.18\arcsec $ \\
LOFAR & 143 & 53 & 3.0 & $ 50.0\arcsec $ \\
WSRT\tablefootmark{a} & 609 & - & 0.7 & $ 48\arcsec \times 28\arcsec $ \\
VLA\tablefootmark{b} & 1400 & 42 & 0.4 & $ 45\arcsec $ \\
\hline
\end{tabular}
\tablefoot{
\tablefoottext{a}{Image provided by K. H. Mack, \citep{Mack1997}}
\tablefoottext{b}{from VLA NVSS}
}
\end{table}

\begin{table}
\centering
\noindent \caption{\small Integrated flux density values.}
\label{table:intflux}
\small
\begin{tabular}{l l}
\hline\hline\\
\small $ \nu $ [MHz] & \small $ S_{\mathrm{int}} $ [mJy] \\
\hline\\
143 & $ 17744 \pm 3568.7 $\\
326 & $ 13132 \pm 140 $\\
609 & $ 8227.7 \pm 90.8 $\\
2695 & $ 3652 \pm 90.8 $\\
4750 & $ 2353.5 \pm 71.2 $\\
10550 & $ 1274.7 \pm 41.7 $\\
\hline
\end{tabular}
\tablefoot{
All other values except the 143 MHz one are taken from \cite{Mack1997}
}
\end{table}

\section{Results}
\label{res}

\subsection{The total intensity structure of 3C~236 at 143 MHz}

\begin{figure*}[!ht]
\centering
\includegraphics[width=\textwidth]{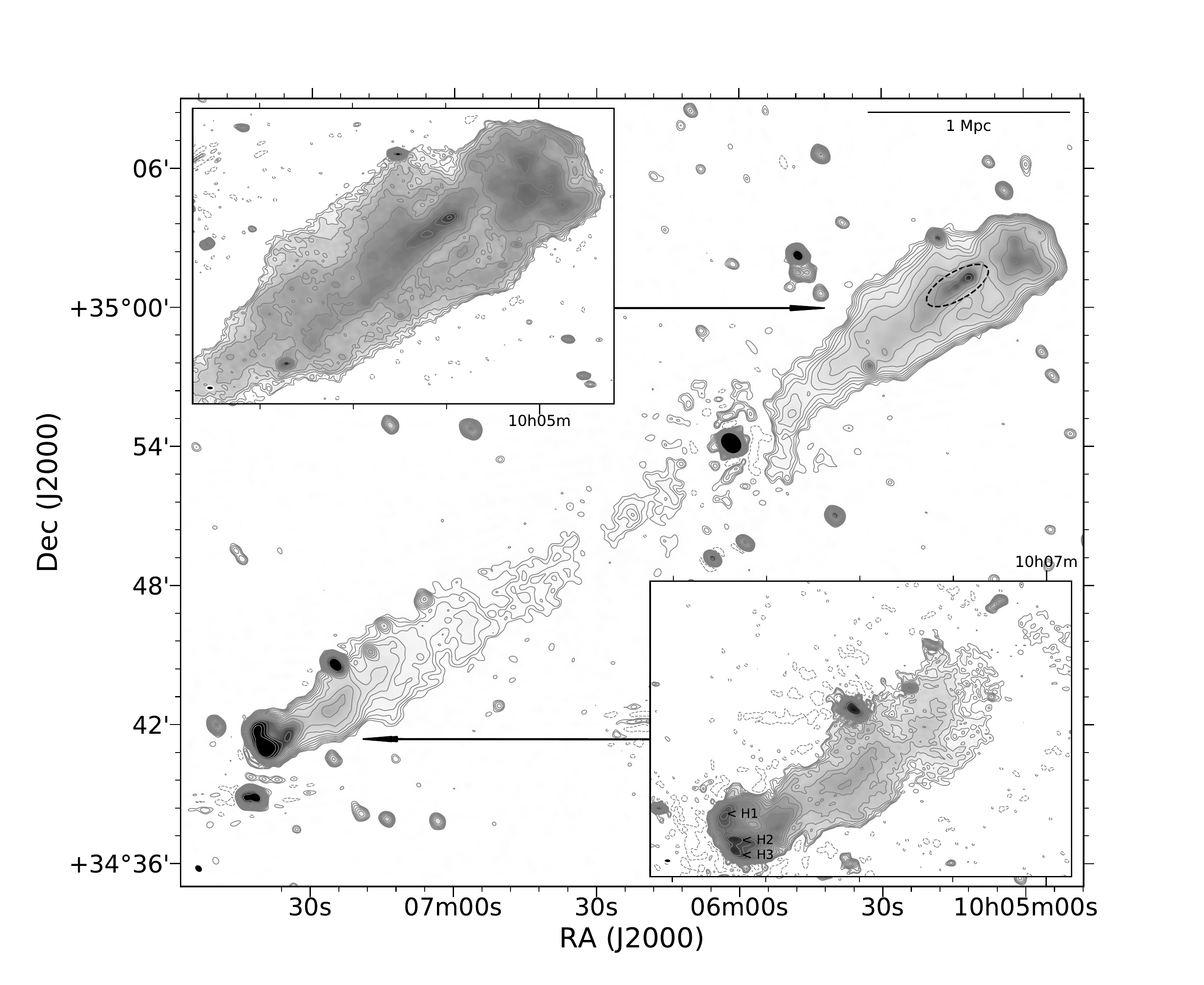}
\caption{
LOFAR intensity map (linear scale, level limits at 1 and 150 \mJybeam) of 3C~236 at 143.6 MHz. Fifteen positive contours are overlaid as solid gray lines with levels at $ \left(\sqrt{2}\right)^{\mathrm{n}} 3\sigma $, where $ \sigma = 0.6 $ \mJybeam , and $ \mathrm{n} $ increasing from $ 0 $ to $ 14 $ in increments of $ 1 $. One negative contour level at $ -3 \sigma $ is overlaid as a dashed gray line. The restoring beam size of $ 23.81\arcsec \times 19.18\arcsec $ is shown in the lower left corner. High resolution image insets (logarithmic intensity scale, limits at 1 , 100 and 500 \mJybeam , $ 11.77\arcsec \times 6.82\arcsec $ restoring beam, $ \sigma = 0.26 $ \mJybeam) of the NW and SE lobes are shown in the top-left and bottom-right corners respectively. Regions of interest are marked with a dashed ellipse and labeled.}
\label{3C236:map}
\end{figure*}

\begin{figure*}[!htpb]
\includegraphics[width=\textwidth]{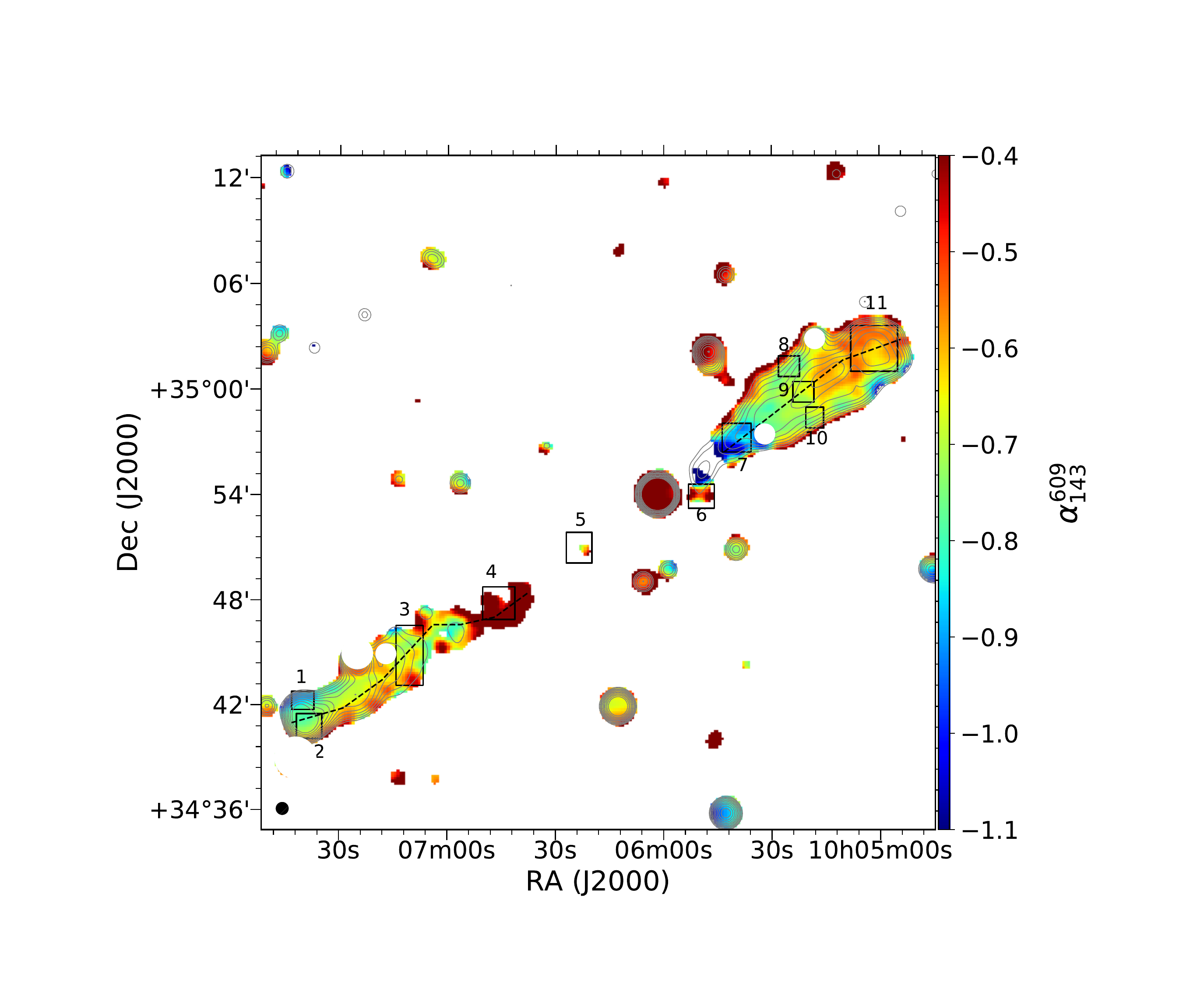}
\caption{$ \alpha_{143}^{609} $ spectral index map. The restoring beam size is $ 50\arcsec $ (bottom left). Overlaid in black are contours tracing the emission from the convolved LOFAR image (listed in Table \ref{table:imgs}) with levels at $ \left(\sqrt{2}\right)^{n} 5\sigma $, where $ n = [0, 9] $ and $ \sigma = 3 $ \mJybeam . Inset are enlarged views of the lobes. Profile paths along which the spectral index values shown in Fig. \ref{3C236:profiles} are measured are shown using dashed lines, as well as measurement regions of the spectral index values listed in Table \ref{table:spix_regs} (solid labelled rectangles). Point sources embedded in the lobes have been masked.}
\label{3C236:spix}
\end{figure*}

The intermediate resolution image ($ 23.8\arcsec \times 19.2\arcsec $) of 3C~236 obtained with LOFAR is shown in the main panel of Fig. \ref{3C236:map}, while the insets zoom in on the two lobes and show their morphology as it appears in our high resolution LOFAR image. Figure \ref{3C236:spix} shows the contour levels of the emission at lower resolution ($ 50\arcsec $), emphasizing better some of the new low surface brightness features detected by LOFAR. An overview of the images is presented in Table \ref{table:imgs}. The map presented in Figure \ref{3C236:map} shows some residual artifacts due to limitation in the calibration around the stronger sources (e.g., the central, very bright compact core of 3C~236), while the regions of diffuse emission are less  affected.

Despite the huge source size (more than half a degree), the LOFAR observations recover the full extent of the source showing, for the first time, its structure at low frequencies and relatively high spatial resolution. The image reproduces well the main (previously known) morphological features of 3C~236 \citep{RefWorks:130}.

The overall structure (about $40\arcmin$ in size, corresponding to $ 4.32 $ Mpc) consists of two very elongated lobes. The north-west (NW) lobe radio emission is more extended transversely to the longitudinal symmetry axis (jet propagation direction) compared to the south-east (SE) lobe (about $ 4\arcmin $ and $ 2\arcmin $ in width towards the NW and SE respectively). At the resolution of the LOFAR observations, the central region including the restarted CSS is unresolved. The asymmetry of the large-scale structure, with the SE lobe extending farther away from the core compared to the NW, is also confirmed by the LOFAR images. Given the size of the source, projection effects are unlikely and hence the source asymmetry must be intrinsic. 

The LOFAR images (especially the one at intermediate resolution) show that both lobes extend all the way to the core. The emission connecting the SE lobe to the core has very low surface brightness (around $ 0.5$ \mJybeam), nevertheless maintaining its width for the full length of the lobe (see the intensity contours in Fig. \ref{3C236:spix}). 
There are no signs of spurs or very extended emission transverse to the lobe extent, with the exception of the extension to the south in the NW lobe, right next to the core (Figure \ref{3C236:spix}). This extension was earlier seen at higher frequencies, although much weaker \citep{RefWorks:130}. It is reminiscent of structures created by back-flows of plasma deposited in the lobe by the jet seen in other \citep[e.g., X-shaped,][]{Leahy1984, Hardcastle2019, Cheung2009, Saripalli2018} radio galaxies.

The high spatial resolution images of the lobes (seen in the insets of Fig. \ref{3C236:map}) show that in the NW lobe, the leading edge does not show a hotspot, but only a diffuse enhancement in surface brightness. However, as first noted by \cite{RefWorks:130}, there is a compact region in the middle of the lobe, confirmed by the LOFAR image (Fig. \ref{3C236:map}). This inner lobe region is split in two clumps (marked by a dashed ellipse in Figure \ref{3C236:map}), the leading one of which is probably a jet termination/hotspot. Structures of this type are seen in other objects \citep[c.f.][]{Malarecki2013}. The location of the hotspot within the lobe suggests that it may trace a separate event of source activity, propagating through the lobe, or tracing an interruption (flicker) in the accretion history of the activity episode that produced the large-scale lobes.
At the end of the SE lobe, a double hot-spot appears to be present (see bottom right inset in Fig. \ref{3C236:map}). For the first time, we observe that the southern hotspot of the pair has itself a double structure, having two distinct brightness peaks (labeled H2 and H3 in the lower right inset of Figure \ref{3C236:map}). This may be a sign of a jet interaction with IGM gas \citep[e.g.][]{Lonsdale1986}, possibly indicating that the jet used to terminate at the location of the H1 hotspot, then the termination point moved to H2 and currently is located at H3. This is consistent with the hypothesis that the most compact hotspot is where the jet terminates at the current epoch \citep{Hardcastle2001}. Also, it would explain why the SE lobe has a steeper spectrum along the northern edge (discussed below). It is also possible that H1 and H2 are created by lateral emission of plasma from the H3 termination shock. In the 3CR sample, one-sixth of the sources have multiple hotspots; the statistics vary depending on the definitions used and source samples under consideration \citep{Valtaoja1984}.

The differences in the structure of the lobes in 3C~236 suggest that they are not only the results of intermittence in the activity and/or changes in the position angle of the jet. Other effects due to e.g. the propagation of the jets in the inner region must have affected the large-scale morphology.

Given the overall size of the source (more than half a degree), several unresolved background sources are embedded in the lobe emission. Their positions are noted in \cite{Mack1997}. Some of these sources are relatively bright, but we find that they do not present an issue for our analysis.

\begin{figure}[!htpb]
\centering
\includegraphics[width=0.5\textwidth]{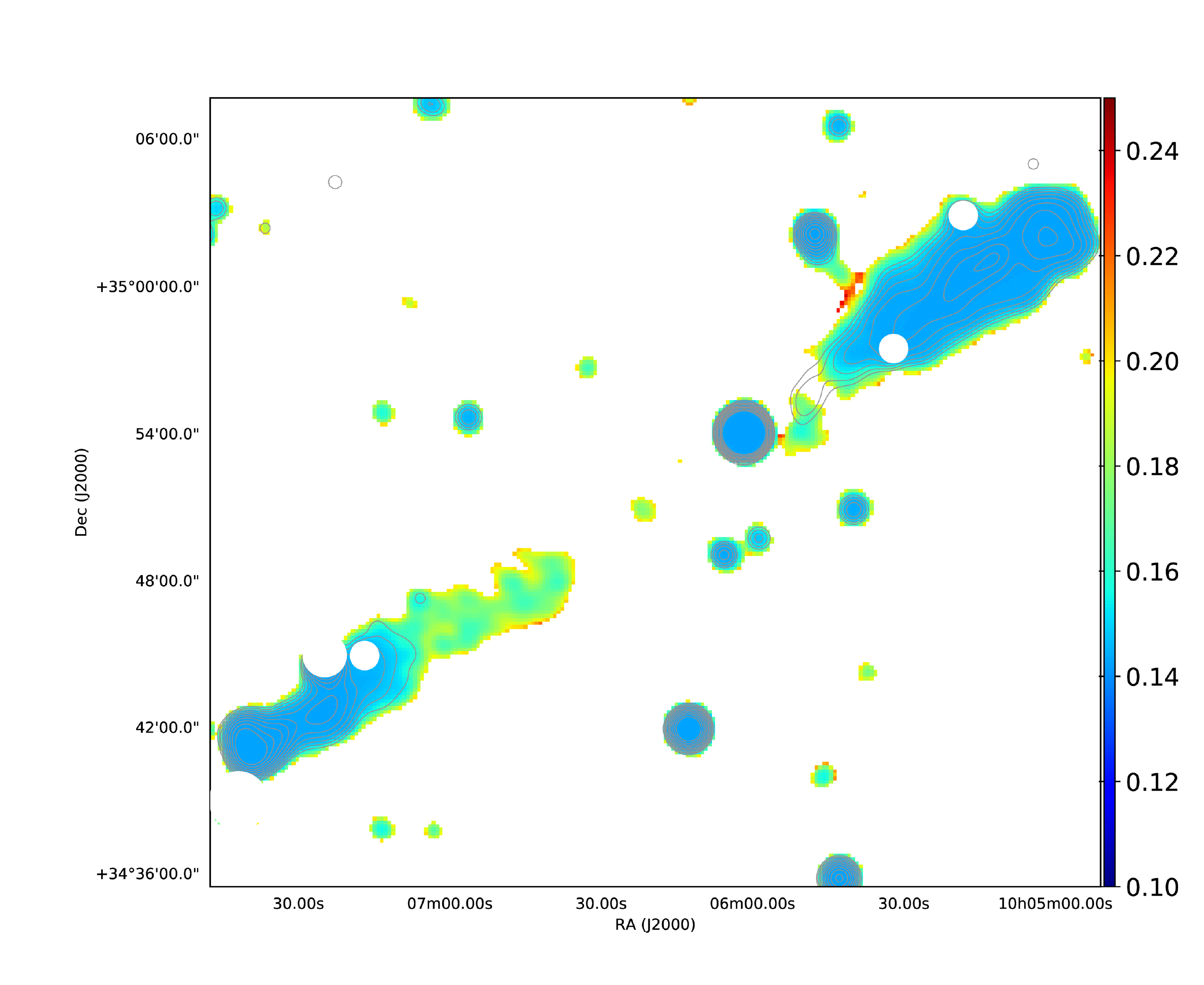}\\
\includegraphics[width=0.5\textwidth]{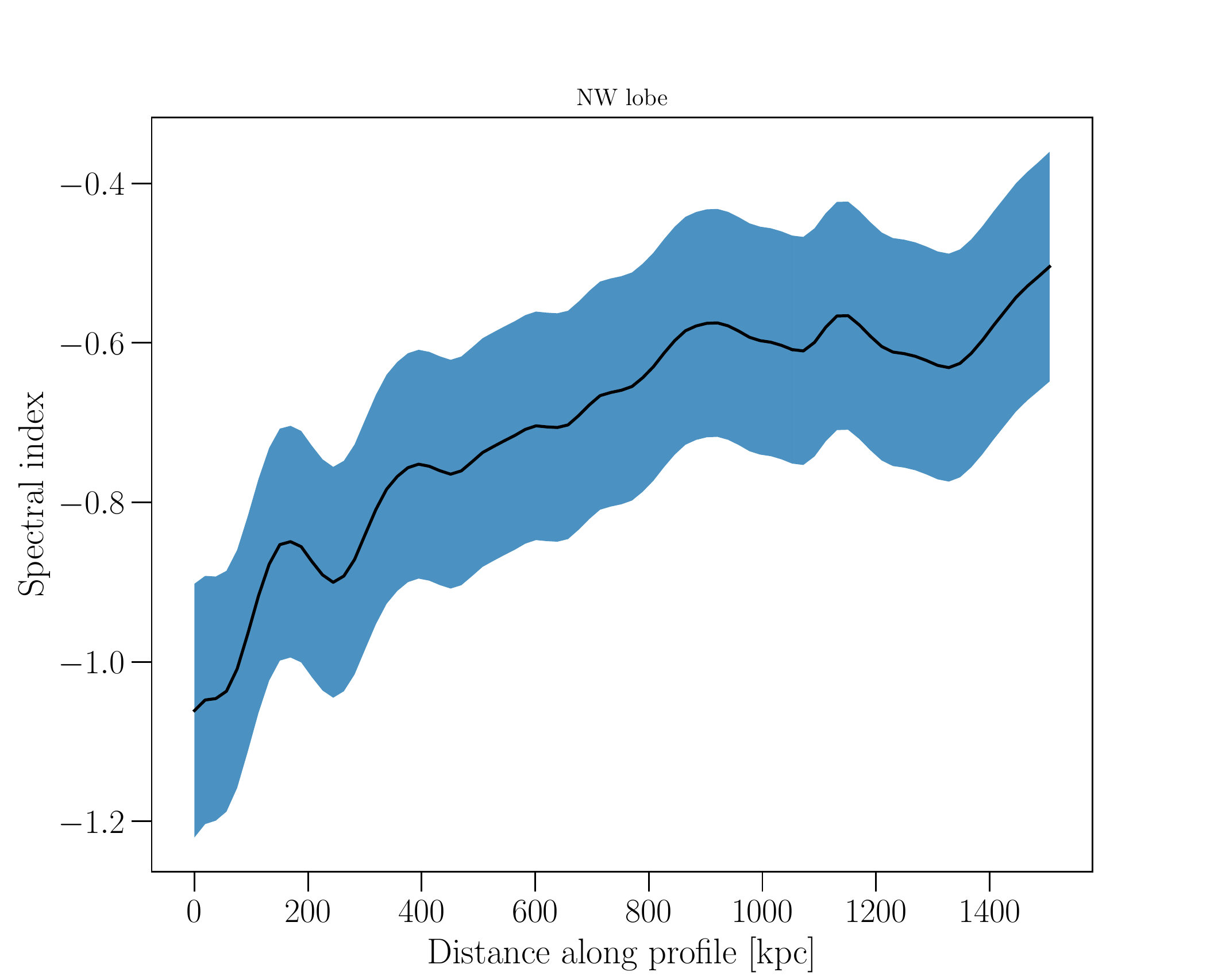}
\includegraphics[width=0.5\textwidth]{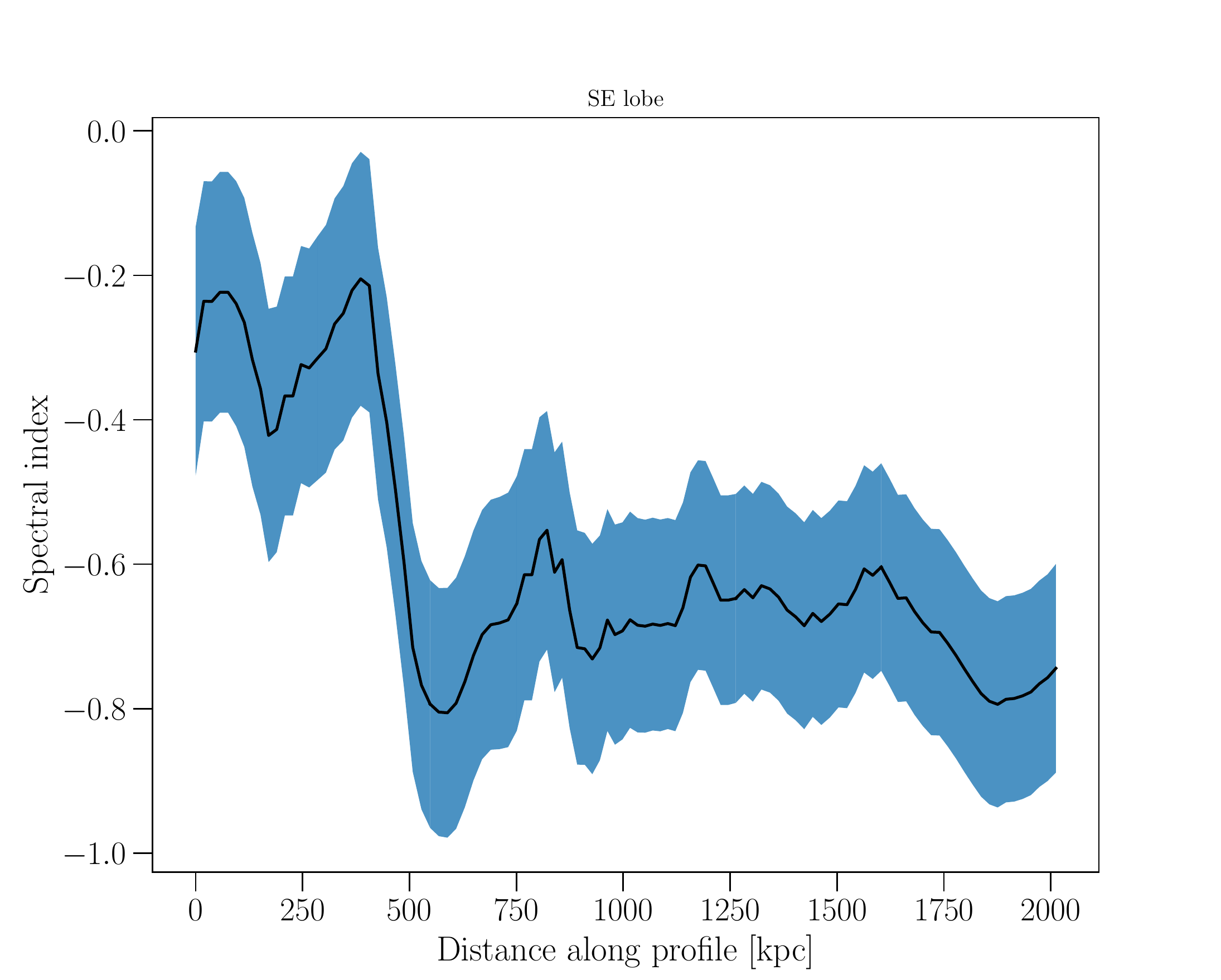}
\caption{Spectral index error (top panel) and spectral index profiles along the paths shown in Fig. \ref{3C236:spix}. The profile paths start in the inner part of the lobes. The shaded area in the spectral profile plots represents the spectral index error.}
\label{3C236:profiles}
\end{figure}

\subsection{Spectral index}
\label{spec_ind}

We have derived a low frequency spectral index ($ S \propto \nu^{\alpha} $) map between 143 and 609~MHz using the images listed in Table \ref{table:imgs}, implementing the following procedure. The 609 MHz image \citep[from][]{Mack1997} was first re-gridded to a J2000 epoch, then we registered the lowest resolution 143~MHz and the 1400~MHz image to the same pixel scale as the 609~MHz image. Finally, we convolved the images with a Gaussian kernel resulting in a circular PSF of $ 50\arcsec $. The re-gridding and convolution were performed using the {\tt imregrid} and {\tt imsmooth} {\tt CASA} \citep{McMullin2007} tasks.

Producing spectral index images from datasets taken with different telescopes is always subject to caveats. In particular, observations at different frequencies must be sensitive to similar spatial structures. The UV-coverage of the datasets can help in checking whether this condition is fulfilled. The UV-range of the mid-resolution LOFAR image (from which the lowest resolution LOFAR map is obtained by convolution) is well matched to that of the WSRT image (upper cut of around $ 7\,\mathrm{k}\lambda $) , so the low frequency spectral index derivation is unaffected by spatial filtering. The NVSS image has relatively few short spacings, i.e., it is less sensitive to extended emission \citep{RefWorks:139,Jamrozy2004}. We keep this limitation in mind when interpreting our (spectral ageing) results.

The flux density scale was taken to be uncertain at a level of 20\% for the LOFAR \citep{Shimwell2017} and 5\% for the WSRT \citep{Mack1997} and VLA \citep{RefWorks:139} observations, respectively. We added these uncertainties in quadrature to the r.m.s. errors of the corresponding maps.

We have derived the $ \alpha_{143}^{609} $ spectral index using the standard expression: $ \alpha = \log(S_{1} / S_{2}) / \log(\nu_{1} / \nu_{2}) $. We show the results in Figure \ref{3C236:spix}, restricted to pixels having flux density values above $5\sigma$ in the corresponding images. The spectral index errors were obtained by error propagation of the flux density errors in the corresponding images:

\[ \delta\alpha = \frac{1}{\ln\frac{\nu_{1}}{\nu_{2}}}\sqrt{\left(\frac{\delta S_{1}}{S_{1}}\right)^{2} + \left(\frac{\delta S_{2}}{S_{2}}\right)^{2}} \]

\noindent here, $ \delta S $ represents the flux density measurement error. The resulting spectral index error map is shown in Fig. \ref{3C236:profiles}, top panel.

We have also measured flux densities in eleven characteristic regions, (along the lobes, encompassing the lobe termination regions and across the NW  lobe). These numbered regions are listed in Table \ref{table:spix_regs}, and indicated in Fig. \ref{3C236:spix}. For each region we have computed the spectral index to investigate whether differences with the values reported in the spectral index map (which samples smaller spatial scales) are present. We find no significant deviations, indicating the robustness of the spectral index map. Also, we show the spectral index profiles along both lobes; the profiles are indicated as dashed lines in Fig. \ref{3C236:spix}.

\begin{table*}[!htpb]
\centering
\noindent \caption{\small Spectral index and spectral ages for the regions defined in Fig. \ref{3C236:spix}.}
\label{table:spix_regs}
\small
\begin{tabular}{c c c c c c c c c}
\hline\hline\\
\small Region ID & \multicolumn{3}{c}{\small $ \alpha_{143}^{609} $} & \small Model & \small $ \alpha_{\mathrm{inj}} $ & \small Spectral age [Myr] & & $ \small \chi^{2}_{\mathrm{red}} $ \\
\hline\\
$ 1 $  & & $ -0.82 \pm 0.14 $ & & JP & $ - $ & $ - $ &  & $ - $ \\
$ 2 $  & & $ -0.74 \pm 0.14 $ & & JP & $ -0.57 $ & $ 51.3^{+7.2}_{-6.9} $ & & $ 0.01 $ \\
$ 3 $  & & $ -0.64 \pm 0.14 $ & & JP & $ -0.20 $ & $ 140.3^{+6.4}_{-1.5} $ & & $ 0.89 $ \\
$ 4 $  & & $ -0.32 \pm 0.14 $ & & JP & $ -0.20 $ & $ 116.7^{+3.5}_{-6.4} $ & & $ 2.63 $ \\
$ 5 $  & & $ -0.60 \pm 0.15 $ & & JP & $ - $ & $ - $ & & $ - $ \\
$ 6 $  & & $ -0.60 \pm 0.15 $ & & JP & $ -0.20 $ & $ 153.2^{+6.3}_{-6.3} $ & & $ 3.56 $ \\
$ 7 $  & & $ -0.88 \pm 0.14 $ & & JP & $ -0.54 $ & $ 159.2^{+2.5}_{-2.6} $ & & $ 0.00 $ \\
$ 8 $  & & $ -0.70 \pm 0.14 $ & & JP & $ -0.20 $ & $ 135.3^{+5.5}_{-1.9} $ & & $ 0.06 $ \\
$ 9 $  & & $ -0.67 \pm 0.14 $ & & JP & $ -0.40 $ & $ 83.8^{+3.5}_{-5.3} $ & & $ 0.03 $ \\
$ 10 $  & & $ -0.68 \pm 0.14 $ & & JP & $ -0.40 $ & $ 91.6^{+2.8}_{-8.0} $ & & $ 0.12 $ \\
$ 11 $  & & $ -0.59 \pm 0.14 $ & & JP & $ -0.40 $ & $ 75.6^{+3.1}_{-9.6} $ & & $ 0.67 $ \\
\hline\\
&\multicolumn{3}{c}{\small $ S_{\mathrm{int}} $[mJy]} & & & & \small $ \nu_{\mathrm{br}} $ [MHz] & \\
& 143[MHz] & 609[MHz] & 1400[MHz] & & & & &\\
\hline\\
\small NW lobe & $ 5030 \pm 80 $ & $ 1930 \pm 390 $ & $ 730 \pm 150 $ & CI & $ -0.55 $ & $ 129.1^{+41.2}_{-30.6} $ & $ 479 $ & $ 0.50 $ \\
\small SE lobe & $ 3580 \pm 100 $ & $ 1260 \pm 250 $ & $ 580 \pm 120 $ & CI & $ -0.55 $ & $ 117.1^{+41.1}_{-30.9} $ & $ 582 $ & $ 0.01 $ \\
\hline\\
\end{tabular}
\end{table*}

The spectral index map shows that the outer lobe regions have spectral index values (between $ -0.5 $ and $ -0.65 $) typical regions with ongoing particle acceleration. This is also observed in the embedded hotspot region in the NW lobe and in the hotspot in the SE lobe (although here the spectral index is around $ -0.1\mathrm{dex} $ steeper). The spectral index generally steepens (see bottom two panels in Figure \ref{3C236:profiles}) toward the edges of the lobes and toward the core (consistent with the FRII overall source morphology), indicating loss of energy and (spectral) ageing of the particle populations as they back-flow in the direction of the core. However, curiously, the SE lobe has a region of very flat spectral index in its core-facing end; a hint of this region is also observed in the higher frequency spectral index maps of \cite{RefWorks:148}. These trends are shown in the spectral index profiles of the lobes presented in Figure \ref{3C236:profiles}.
The SE lobe has the flattest spectral index along its southern edge. There is a transition to steeper spectral index values of $ \sim -0.9 $ along the northern edge of its outer region. Whereas the interpretation of the spectral index map in this region is not straightforward, the observed steepening could be real at least in some parts and warrants further investigation in future studies of this object.

\cite{RefWorks:148} derive $ \alpha_{326}^{609} $ spectral indices for the NW lobe of around $ -1 $ for the outer regions to $ -1.2 $ going toward the core. Our $ \alpha_{143}^{608} $ spectral index map shows much flatter spectral index values, around $ -0.5 $ to $ -0.6 $, (c.f. the spectral profiles in Figure \ref{3C236:profiles}) meaning that LOFAR detects the particle population related to the primary injection of energy in the lobe. For the SE lobe, the agreement between our spectral index values and those derived by \cite{RefWorks:148} is better, although we have flatter values, with $ \alpha_{143}^{609} \sim -0.6 $ versus their values of $ \alpha_{326}^{609} \sim -0.8 $ for the outer lobe. We have derived the spectral index for several regions in the source (Table \ref{table:spix_regs}) to test whether the values we obtain in our spectral index map are reliable; the spectral index values per region match those from the map. High resolution mapping helps to disentangle the detailed spectral behaviour.

\subsection{Source energetics and radiative ages}

Before discussing the 3C~236 energetics, we estimate the magnetic field value throughout the source. We make the assumption that the field structure in the lobes is uniform and oriented perpendicular to the line of sight. We use cylindrical geometry for the lobes, and calculate the magnetic field strength assuming that equal energy is contained in the relativistic particles and the field (equipartition assumption). Furthermore, we set the ratio of electrons to protons to unity, as well as the filling factor of the lobes. We adopt limits of the spectral range from 10~MHz to $ 10^{5} $ MHz, and we set the spectral index of the radiation in this range to $ \alpha = -0.85 $ (taking into account the observed values at low frequencies, as well as assuming spectral steepening to higher frequencies). Using the relation by \cite{Miley1980}, calculating at a frequency of 143 MHz (Table \ref{table:fluxes}) and averaging over both lobes, we obtain $ \mathrm{B} = 0.75 \, \mathrm{\mu} $G for the equipartition magnetic field strength. As was noted by \cite{Brunetti1997}, the equipartition field calculated in this manner should be corrected, to take into account a low energy cut-off value for the spectrum of the particles and a value for the particle spectral index at injection time. With $ \gamma_{min} = 200 $, and $ \mathrm{\alpha}_{\mathrm{inj}} = -0.75 $ (average low frequency spectral index in the lobes), we find $ \mathrm{B} = 1.28 \, \mathrm{\mu} $G for the average source magnetic field, a value we will be using further in our analysis. This value of the magnetic field is lower than the CMB equivalent magnetic field at the source redshift ($ B_{CMB} = 3.93 \, \mathrm{\mu} $G). Thus, the dominant energy loss mechanism of the relativistic particles generating the synchrotron radio emission will be inverse Compton scattering off the omnipresent CMB photons.

\begin{figure}[!htpb]
\centering
\includegraphics[width=0.5\textwidth]{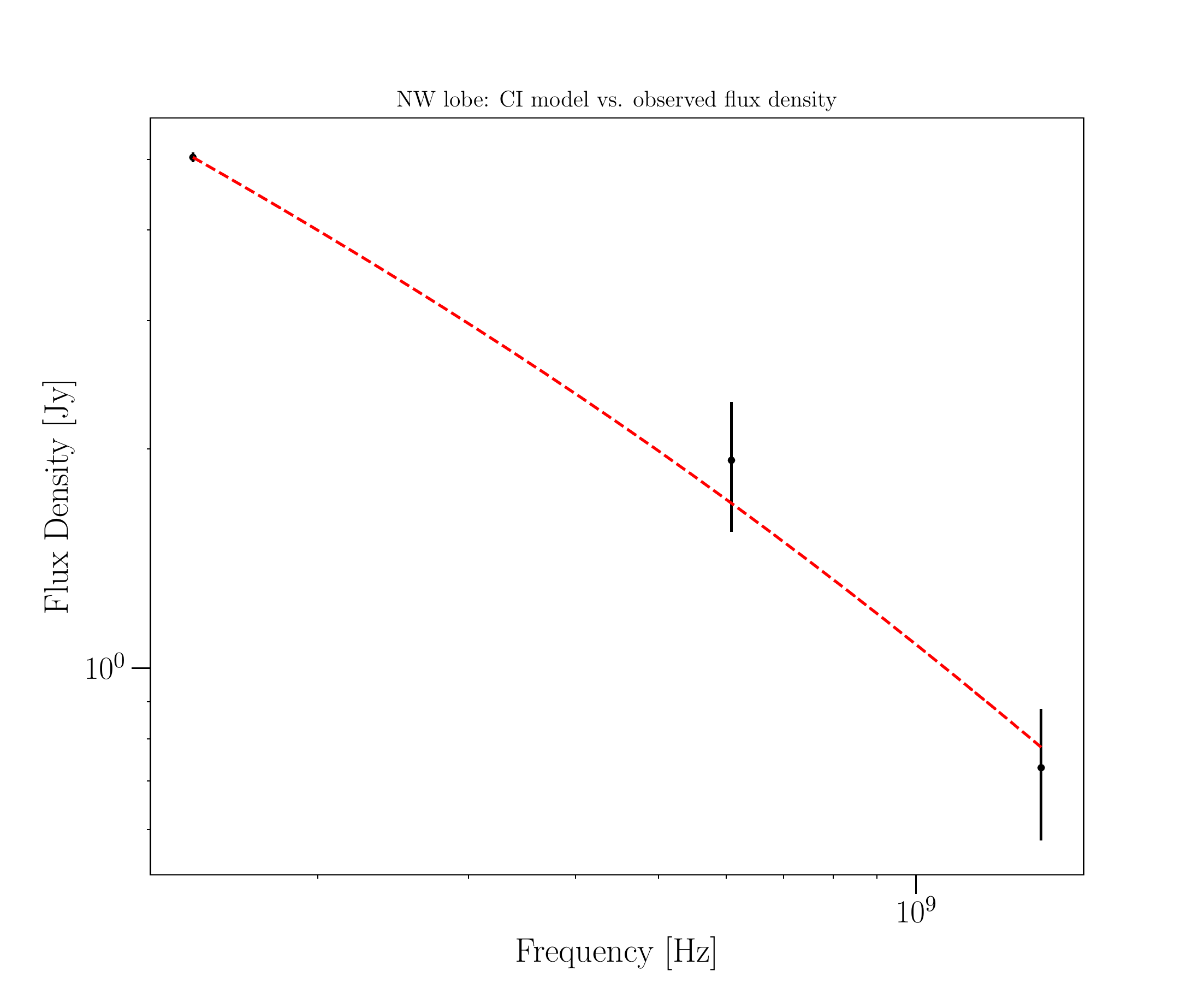}\\
\includegraphics[width=0.5\textwidth]{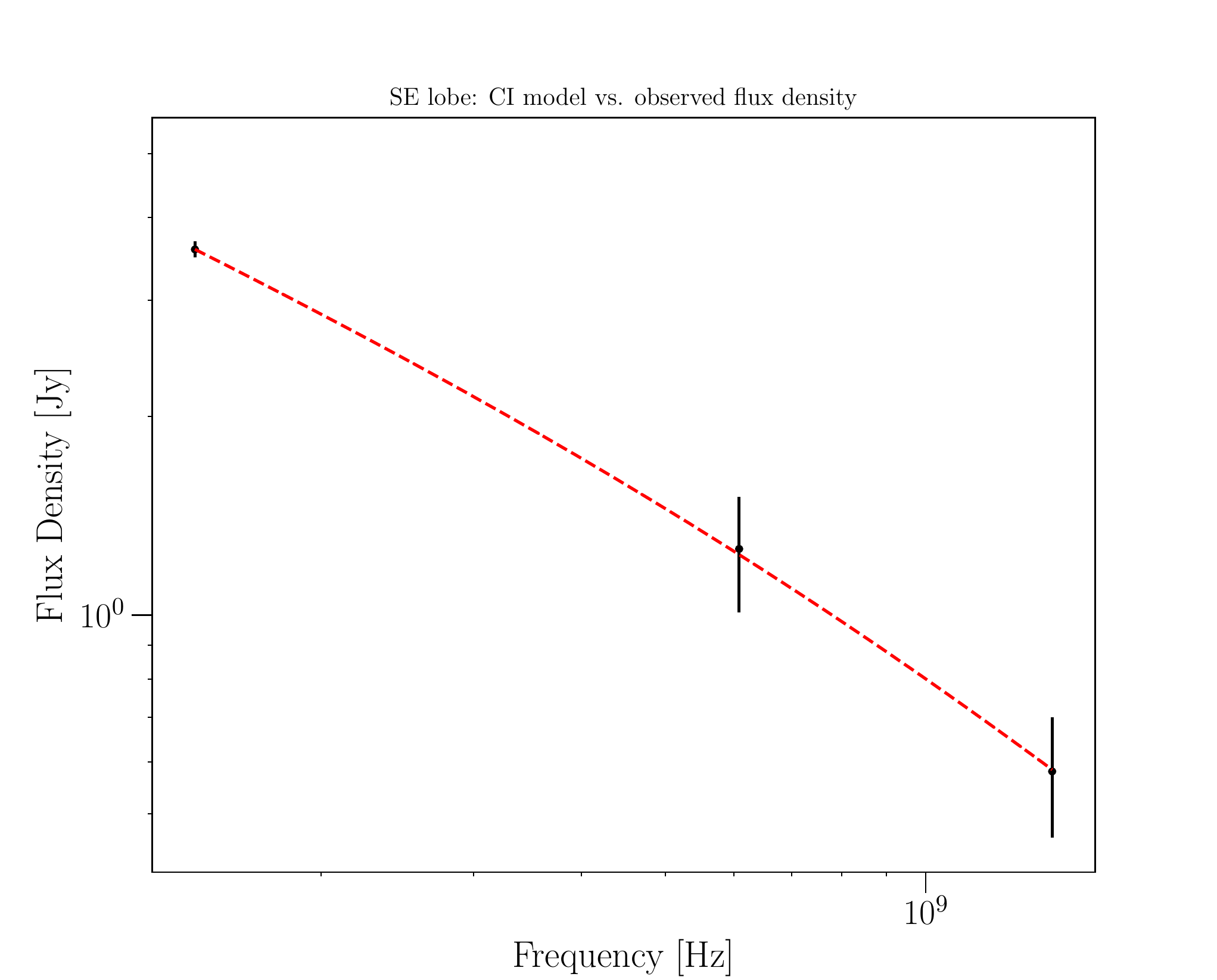}
\caption{CI model fits as reported in the lower section of Table \ref{table:spix_regs}.}
\label{CI_fits}
\end{figure}

The spectral ages of the emitting regions are calculated using two different approaches: first for the regions defined in Figure \ref{3C236:spix} and second for measurement regions encompassing the NW and SE lobes separately, avoiding embedded point sources and measuring out to the $ 5\sigma $ contour in the 143 MHz image. We have used the {\tt fitintegrated} and {\tt fitcimodel} tasks of the {\tt BRATS}\footnote{http://www.askanastronomer.co.uk/brats/} software package \citep{Harwood2013,Harwood2015} for the two cases respectively. The fitting was performed using flux density measurements at three different frequencies, using the low resolution LOFAR image and the WSRT and VLA images listed in Table \ref{table:imgs}. In the {\tt fitintegrated} task we fitted a Jaffe-Perola \citep[JP,][]{Jaffe1973} model, and the {\tt fitcimodel} task fits a continuous injection (CI) model to the integrated flux densities for the source regions under consideration. The CI model was used when modelling the lobes assuming that they encompass regions where particle acceleration is ongoing, which can not be stated for (all) of the smaller measurement regions. Although the models do not give intrinsic ages \citep{Harwood2017}, they are useful to address the source properties. The injection spectral index was kept fixed for each fitting run, and the fitting was performed over a search grid spanning values from $ \alpha_{\mathrm{inj}} = -0.2 $ to $ \alpha_{\mathrm{inj}} = -0.9 $. The derived spectral ages and spectral break frequencies (in the CI fit case) resulting from the fitting procedure are shown in Table \ref{table:spix_regs}. The average reduced $ \chi^{2} $ measure for the goodness of the fit (one degree of freedom) is given in the last column. The fit results are shown in Figure \ref{CI_fits}.

The derived ages for individual regions indicate older plasma as one goes from the lobe outer edges toward the core, consistent with what would be expected if the main acceleration regions are located in the outer lobes. The injection spectral indices which best fit the data are not steep, indicating that LOFAR is probing particles which have retained their energy since their acceleration phase. We will discuss the spectral ages further in Section \ref{dis}.

The particle energy assuming equipartition can be expressed as \citep{RefWorks:148,Pacholzcyk1970}

\[ \mathrm{E}_{\mathrm{eq}} = \dfrac{7}{4}\left[(1+\mathrm{k})\mathrm{c}_{\mathrm{12}}\mathrm{P}\right]^{\frac{4}{7}}\left(\dfrac{\mathrm{\Phi} \mathrm{V}}{\mathrm{6\pi}}\right)^{\frac{3}{7}} \]

\noindent where $ \mathrm{\Phi} = 1 $ is the volume filling factor, $ \mathrm{V} $ the volume of the region filled with relativistic particles and fields, $ \mathrm{k} $ (= 1) the electron to proton ratio, $ \mathrm{P} $ the radio luminosity integrated over the earlier specified frequency range, for the regions under consideration, and $ \mathrm{c}_{\mathrm{12}} $ is a constant \citep[in our case $ \mathrm{c}_{\mathrm{12}} = 1.6 \times 10^{7} $;][]{Pacholzcyk1970}.

Assuming that the lobes are in pressure balance with the intergalactic medium (IGM), the relativistic gas pressure in the lobes, $ \mathrm{p}_{\mathrm{l}} $\footnote{$ \mathrm{p}_{\mathrm{l}} = (\gamma - 1)\mathrm{e}_{\mathrm{eq}} $, where $ \mathrm{e}_{\mathrm{eq}} = \mathrm{E}_{\mathrm{eq}} / \mathrm{V} $ is the lobe energy density and $ \gamma $ is the ratio of specific heats; for relativistic gas $ \gamma = \frac{4}{3} $} should balance with the gas pressure of the IGM ($ \mathrm{p}_{\mathrm{IGM}} = \mathrm{n}_{\mathrm{IGM}}\mathrm{kT} $), where $ \mathrm{k} $ is the Boltzmann constant and $ \mathrm{T} $ the temperature of the IGM in degrees Kelvin. Adopting $ \mathrm{T} = 10^{7} $K, we can roughly estimate the IGM particle density values $ \mathrm{n}_{\mathrm{eq}} = \frac{\mathrm{e}_{\mathrm{eq}}}{3\mathrm{kT}} $ \citep{RefWorks:148, Hunik2016}, and list the resulting  values in Table \ref{table:fluxes}.

\begin{table*}[!htpb]
\centering
\noindent \caption{\small Source parameters and derived quantities}
\label{table:fluxes}
\small
\begin{tabular}{c c c c c c c}
\hline\hline\\
\small ID & \small $\mathrm{S}_{\mathrm{143}}$ [Jy] & \small $\mathrm{L}_{\mathrm{143}}$ [W$\mathrm{Hz}^{-1}$] & $ \mathrm{E}_{\mathrm{eq}} $ [J] & $ \mathrm{V} $ [$ \mathrm{m}^{3} $] & $ \mathrm{n}_{\mathrm{IGM}} $ [cm$ ^{-3} $] & $ \mathrm{P} $ [Pa] \\
\hline
NW lobe & $ 5.0 $ & $ 1.2 \times 10^{26} $ & $ 2.0\times10^{53} $ & $ 0.9\times10^{67} $ & $ 5.2 \times 10^{-5} $ & $ 7.2\times10^{-15} $ \\
Core & $ 9.0 $ & $ 2.2 \times 10^{26} $ & - & - & - & - \\
SE lobe & $ 3.6 $ & $ 8.9 \times 10^{25} $ & $ 1.7\times10^{53} $ & $ 1.0\times10^{67} $ & $ 4.2\times10^{-5} $ & $ 5.8\times10^{-15} $ \\ 
\hline\\
\end{tabular}
\end{table*}

\section{Discussion}
\label{dis}

Our LOFAR imaging recovers the source structure as described previously in the literature \citep{RefWorks:130, Mack1997}, and discussed in the previous section of this work. Owing to the high surface brightness sensitivity in our LOFAR images, we can now trace the SE lobe emission all the way to the core even in the intermediate resolution maps. We note that the NW lobe is shorter than the SE lobe. Interestingly, this asymmetry is inverted for the small-scale emission of the CSS core. There, the NW extension is longer than the SE as seen by \cite{RefWorks:258}. They also speculate that the dust lane imaged by HST close to the core may be part of the material that helps collimate the radio emission.

As was noted in Section \ref{intro}, there is a hint of a slight wiggle of the ridge line connecting the core and the outer lobe edges. It is visible in Figure \ref{3C236:map} as a departure from the symmetry axis in the NW lobe, where the inner hotspot and the outer diffuse region are not on a straight line to the core. This may be due to the wobble of the jet as it drills through the IGM/ICM over time. In this context, the appearance of the SE lobe hotspot is intriguing. It was described as a double hotspot in the literature \citep{RefWorks:130}; now, using LOFAR, we can see that it is in fact a triple hotspot; the southern component is split in two (Figure \ref{3C236:map}). It may be that the jet was deflected at the hotspot producing the observed morphology, or that the jet working surface has shifted over time. \cite{Lonsdale1986} have suggested that such hotspots can originate from a flow redirection in the primary hotspot.

\cite{RefWorks:228} have classified 3C~236 as a double-double \citep{Saikia2009,Orru2015} radio galaxy, since the restarted activity in the core has extended emission aligned with the large-scale lobes. Then, 3C~236 may be a "triple-double" radio galaxy; the inner hotspot in the NW lobe signifying a stall in the jet, or a short sputter in the accretion episode responsible for the creation of the large-scale lobes. In this view, the diffuse outer region of the NW lobe is the (still active) remnant hotspot of the previous activity episode and the embedded hotspot is produced by the jet expelled during the re-activation episode, which is still advancing through the lobe material. Within this context, the wiggle noticeable in the source morphology (mentioned above) can be explained by a slight shift in the jet axis during the jet stall/sputter event.

The lobe length asymmetry and the position of the hotspots may be caused by a difference in material density on the opposite sides of the host galaxy, at the position of the lobes, and higher for the NW lobe. This is tentatively supported by the particle density we have derived, presented in Table \ref{table:fluxes}, which is $ \sim 3 $ times higher than the medium density obtained by \cite{Mack1997} and hence broadly comparable with their result.

Owing to their sizes, GRGs can be used as probes of the physical conditions of the IGM. \cite{Malarecki2013} have performed such a study on a sample of 19 GRGs; we are in agreement with the values they have derived for the mean lobe pressures in their sample (ranging from $ 1.34\times10^{-15} $ to $ 1.91\times10^{-14} $ Pa). In a subsequent study of the same sample \citet{Malarecki2015} find that GRGs tend to occur in sparse environments (such as the one of the 3C~236 host galaxy), and they show tentatively that shorter lobes expand in regions of (on average) higher galaxy density. This may be relevant to explain the lobe morphology of 3C~236. Further studies on the immediate environment of the host galaxy of 3C~236 should test this hypothesis. If true, the environment should be denser to the north-west, where the shorter lobe extends. On the other hand, the large-scale asymmetry and the (reverse) small-scale asymmetry may have a physical origin in asymmetric host galaxy properties.

Recent studies of the GRG NGC~6251 by Cantwell et al. (submitted) have found ages for its lobes of less than 50  Myr, and show that the newly discovered faint lobe extensions have ages greater than 200 Myr. The radiative ages for the lobes of 3C~236 we derive fall in between of the ages derived for the different regions of NGC~6251. Given the morphological difference between the lobes of these two GRGs (the lobes of NGC~6251 are far less confined, and it is an FRI radio galaxy), the results of the studies are consistent.  The lobe pressure values they find ($ 4.9\times10^{-16} $ to $ 4.8\times10^{-13} $ Pa) are also consistent with our findings.
Our derivations of the lobe ambient medium assume that the lobes are in pressure balance with the IGM. One may find this assumption objectionable, as radio galaxy lobes are often observed to be under-pressured. However, \cite{Croston2014} have argued that FRIs can be in pressure balance if there is entrainment of surrounding material by the jet flow. Recent reports that FRI lobes are energetically dominated by protons \cite{Croston2018} seem to support the entrainment scenario. Similarly, \cite{harwood2016, harwood2017b} argue that FRIIs can be brought back into agreement by considering the steeper than expected injection index which is sometimes derived using model fits to data from low-frequency observations.

Our 3C~236 spectral index map, with the highest spatial resolution obtained so far at these frequencies, allows us for the first time to associate morphological and spectral features. We clearly see a flatter (compared to the surrounding regions) spectral index associated with the inner hotspot of the NW lobe, hinting at it being a particle acceleration region. Also, while previous spectral index maps (Figure 3, top panel in \cite{RefWorks:148}) only weakly hinted at the spectral index steepening toward the lobe edges, we can now better trace that transition. The curious flattening of the spectral index in the inner SE lobe which was hinted at previously \citep{RefWorks:148}, now stands out. It can be a signature of the interaction between the jet inflating the SE lobe and the IGM at that position; the spectral index indicating that acceleration is ongoing.

In general, the spectral ages we have derived do not agree with the values published by \cite{RefWorks:148}; these authors obtain lower age estimates (ranging from less than $ 8 $ Myr to $ 20 $ Myr). The only exception is region two, where our age is broadly comparable with their estimate (they have an age of around $ 20 $ Myr for that region of the source, while our value is around $ 50 $ Myr, both using the JP model). The ages we derive measuring the integrated flux density of the lobes are substantially higher than what \cite{RefWorks:148} derive. The fact that the age for region eight using the single-injection JP model is comparable with the age for the NW lobe derived using the CI model, suggests that our ages are a more robust measure of the characteristic age compared to previous studies. The break-frequencies in our model lobe spectra are located in the lower frequency range of the data used by \cite{RefWorks:148}, which is consistent with the fact that we measure flatter spectral indices in the lobes. LOFAR can trace the spectral flattening towards still lower frequencies, and thus characterize the spectral break.

The age difference between these studies is most likely due to the fact that LOFAR measures the emission from the oldest particle population, affecting our estimates. It should also be noted that due to the uncertainties in the assumed values of the input parameters (especially the magnetic field value; \cite{RefWorks:148} use values between 0.3 $ \mu $G and 0.7 $ \mu $G), uncertainties in the model, mapping resolution as well as the sparse frequency coverage, these ages should be taken as limits to the actual values.

Our age estimates support a scenario where the lobes are built up by a jet advancing with a speed of around $ 0.1\mathrm{c} $ \citep[as argued by][]{RefWorks:130}, i.e. that speed is required to inflate the lobes to their present linear size in a time broadly consistent with their derived ages (overall ages based on the CI model). Further, as was already mentioned in the introduction section of this work, \cite{RefWorks:228} suggest (based on their HST studies of star formation in the nucleus of the host galaxy) an age of the large-scale lobes in the range of $ 10^{8} $ to $ 10^{9} $ years, in line with our findings.

\section{Conclusion}
\label{con}

We have presented new LOFAR observations of the GRG 3C~236. We have studied this radio galaxy for the first time at a resolution of up to $ 6\arcsec $ at 143 MHz. Also, we have derived the highest resolution spectral index maps to date (at $ 50 \arcsec $ resolution). Our main conclusions are:

\begin{itemize}
\item We observe an inner hotspot in the north-western lobe, separate from its more diffuse outer region. It is also discernible in the spectral index map, as a region undergoing more recent particle acceleration (flatter spectral index values). This detection, taken together with the overall source morphology, may be an indication of a short interruption of the accretion episode/jet sputter.
\item The brighter component of the SE lobe double hotspot is resolved in two components, making this feature a triple hotspot.
\item The source energy / pressure balance with the IGM suggests that confinement by the IGM may be responsible for the morphology of the lobes; the NW lobe  is probably confined and the SE one has expanded in a lower density medium, reflected in the somewhat steeper spectrum of its outer region / northern edge.
\item The derived spectral ages are consistent with a jet advancing at $ 0.1 $c in the surrounding medium of the host galaxy.
\end{itemize}

LOFAR is a valuable instrument for studies of giant radio sources. Its sensitivity to low surface brightness features, combined with its capability for high resolution imaging at low frequencies, offers an unprecedented detailed view of source emission regions containing low energy plasma. This is useful to uncover previously unknown features even in targets which have been studied for decades, such as 3C~236.

\begin{acknowledgements}

LOFAR, the Low Frequency Array designed and constructed by ASTRON, has facilities in several countries, that are owned by various parties (each with their own funding sources), and that are collectively operated by the International LOFAR Telescope (ILT) foundation under a joint scientific policy.
We would like to thank Karl-Heinz Mack for providing fits images for the previously published WSRT map.
RM gratefully acknowledges support from the European Research Council under the European Union's Seventh Framework Programme (FP/2007-2013) / ERC Advanced Grant RADIOLIFE-320745.
MB acknowledges support from INAF under PRIN SKA/CTA ‘FORECaST’
GJW gratefully acknowledges support from the Leverhulme Trust.
SM acknowledges funding through the Irish Research Council New Foundations scheme and the Irish Research Council Postgraduate Scholarship scheme.
This research has made use of the NASA/IPAC Extragalactic Database (NED), which is operated by the Jet Propulsion Laboratory, California Institute of Technology, under contract with the National Aeronautics and Space Administration.\\
This research has made use of APLpy, an open-source plotting package for Python hosted at http://aplpy.github.com
This research made use of Astropy, a community-developed core Python package for Astronomy (Astropy Collaboration, 2019).

\end{acknowledgements}

\bibliographystyle{3C236_lofar.bst}
\bibliography{3C236_lofar.bib}

\end{document}